\documentclass{aa}
  \usepackage{txfonts}
 \usepackage{graphicx}
 
\begin{document}

\title{Out of the darkness: the infrared afterglow of the \textit{INTEGRAL}\thanks{Based on observations with \textit{INTEGRAL}, an ESA project with instruments 
and science data centre funded by ESA member states (especially the PI 
countries: Denmark, France, Germany, Italy, Switzerland, Spain), Czech Republic and Poland, and with the participation of Russia and the USA. } burst \object{GRB\,040422} observed with the VLT\thanks{Based on observations made with ESO telescopes at the Paranal Observatory under programme 073.D-0255.}}
\author{P. Filliatre\inst{1,2}
	\and P. D'Avanzo\inst{3,4}
	\and S. Covino\inst{5}
	\and D. Malesani\inst{6}
	\and G. Tagliaferri\inst{5}
	\and S. McGlynn\inst{7}
	\and L. Moran\inst{8}
	\and P. Goldoni\inst{2,1}
	\and S. Campana\inst{5}
	\and G. Chincarini\inst{5,4}
	\and L. Stella\inst{9}
	\and M. Della Valle\inst{10}
	\and N. Gehrels\inst{11}
	\and S. McBreen\inst{12}
	\and L. Hanlon\inst{7}
	\and B. McBreen\inst{7}
	\and J. A. Nousek\inst{13}
	\and R. Perna\inst{14}}
\offprints{P. Filliatre, \email{filliatr@cea.fr}}
	
\institute{Laboratoire Astroparticule et Cosmologie, UMR 7164, 11 place Marcelin Berthelot, F-75231 Paris Cedex 05, France
	\and Service d'Astrophysique, CEA/DSM/DAPNIA/SAp, CE-Saclay, Orme des Merisiers, B\^{a}t. 709, F-91191 Gif-sur-Yvette Cedex, France	
	\and Universit\`{a} dell'Insubria, Dipartimento di Fisica e Matematica, via Vallegio, 11, I-22100 Como, Italy
	\and Universit\`{a} degli studi di Milano-Bicocca, Dipartimento di Fisica, Piazza delle Scienze 3, I-20126 Milano, Italy
	\and INAF, Osservatorio Astronomico di Brera, via E. Bianchi 46, I-23807 Merate (LC), Italy
	\and International school for advanced studies (SISSA/ISAS), via Beirut 2-4, I-34014 Trieste, Italy
	\and Department of Experimental Physics, University College, Dublin 4, Ireland
	\and Department of Physics \& Astronomy, University of Southampton, Southampton, SO17 1BJ, United Kingdom
	\and INAF, Osservatorio Astronomico di Roma, Via Frascati 33, Monteporzio Catone, I-00040 Rome, Italy
	\and INAF - Osservatorio Astrofisico di Arcetri, largo E. Fermi 5, 50125  Firenze, Italy
	\and NASA Goddard Space Flight Center, Code 661, Greenbelt, MD 20771
	\and Astrophysics Missions Division, Research Scientific Support Department of ESA, ESTEC, Noordwijk, The Netherlands
	\and Department of Astronomy and Astrophysics, Pennsylvania State University, 525 Davey Laboratory, University Park, PA 16802
	\and Department of Astrophysical and Planetary Sciences, University of Colorado at Boulder, 440 UCB, Boulder, CO, 80309, USA}

\date{Received <date> / Accepted <date>}

\titlerunning{The afterglow of the \object{GRB\,040422}}
\authorrunning{P. Filliatre et al.}

\abstract{\object{GRB\,040422} was detected by the \textit{INTEGRAL} satellite at an angle of only 3 degrees from the Galactic plane. Analysis of the prompt emission observed with the SPI and IBIS instruments on \textit{INTEGRAL} are presented.
 The IBIS spectrum is well fit by the Band model with a break 
energy of $E_{0}=56\pm2\,\rm keV$ and $E_{\rm peak}=41\pm3\,\rm keV$. The peak flux is $1.8\times10^{-7}\,\rm erg\,cm^{-2}\,s^{-1}$ and fluence $3.4\times10^{-7}\,\rm erg\,cm^{-2}$ in the range 20--200~keV. 
We then present the observations of the afterglow of \object{GRB\,040422}, obtained with the ISAAC and FORS\,2 instruments at the VLT less than 2 hours after the burst. We report the discovery of its near-infrared afterglow, for which we give here the astrometry and photometry. No detection could have been obtained in the $R$ and $I$ bands, partly due to the large extinction in the Milky Way. We re-imaged the position of the afterglow two months later in the $K_{\rm s}$ band, and detected a likely bright host galaxy. We compare the magnitude of the afterglow with a those of a compilation of promptly observed counterparts of previous GRBs, and show that the afterglow of \object{GRB\,040422} lies at the very faint end of the distribution, brighter only than that of \object{GRB\,021211}, singled out later and in the optical bands, and \object{GRB\,040924} after accounting for Milky Way extinction. This observation suggests that the proportion of dark GRBs can be significantly lowered by a more systematic use of 8-m class telescopes in the infrared in the very early hours after the burst. 
}

\maketitle

\section{Introduction}

The study of GRB afterglows is a promising tool for cosmology, as their absorption spectra convey information on the distance and the chemical environment of a new set of galaxies (e.~g. \cite{fiore}), with the possibility of exploration up to the reionization epoch (\cite{lamb}). However, a debated fraction of GRBs (from less than 10\% (\cite{hete}) to 60\% (\cite{lazzati})) did not show any detectable afterglow in the optical band. Popular and non-mutually exclusive explanations are: these bursts have intrinsically faint afterglows in the optical band (e.~g. \cite{fynbo}; \cite{lazzati}); their decay is very fast (\cite{berger}); the optical afterglow is extinguished by dust in the vicinity of the GRB or in the star-forming region in which the GRB occurs (e.~g. \cite{lamb2000}; \cite{reichart}); their redshift is above 6, so that the Lyman-$\alpha$ absorption by neutral hydrogen in the host galaxy and along the line of sight damps the optical radiation of the afterglow (\cite{lamb2000}). To these physical explanations, one must add the possibility that the search techniques are not accurate and quick enough (\cite{hete}). The possibility that some afterglows are intrinsically faint has of course the most important impact on the modelling of the GRBs as well as their application in cosmology. On the other hand, if one or several of the other explanations are correct, a substantial reduction of the fraction of dark bursts can be achieved by quick observations in the infrared.\\
For this reason, we have adopted a strategy to promptly search for GRB counterparts both in the near-infrared (NIR) and in the optical. To date, several attempts were performed, sometimes leading to a detection (like for \object{GRB\,040827}: \cite{gcn2685}) and sometimes just to upper limits (like for \object{GRB\,040223}: \cite{gcn2535}). This paper deals with the observations of \object{GRB\,040422}, performed with the ISAAC and FORS\,2 instruments installed at the focal plane of the Very Large Telescope (VLT) at the European Southern Observatory (ESO). In this case, a positive detection of the afterglow was nicely achieved.\\
ESA's International Gamma-Ray Astrophysics Laboratory \textit{INTEGRAL} 
(\cite{winkler}), launched in October 2002, 
is composed of two main telescopes, an imager IBIS 
(Imager on Board the \textit{INTEGRAL} Satellite, \cite{uber2003})
 and a spectrometer SPI (Spectrometer on \textit{INTEGRAL}, \cite{ved2003})
coupled with two monitors, one in the X-ray band and in one the optical band. Although not built as a dedicated GRB-mission, 
\textit{INTEGRAL} has a burst alert system called IBAS (\textit{INTEGRAL} Burst Alert System, \cite{Mere03}). 
IBAS carries out rapid localizations for GRBs incident on the IBIS detector 
with a precision of a few arcminutes (\cite{mere:2004}). The public distribution of these coordinates enables multi-wavelength searches for  afterglows at 
lower energies. \textit{INTEGRAL} data on the prompt emission in
 combination with the early multi-wavelength studies, such as presented in this work, can probe these high energy phenomena.
\object{GRB\,040222} was detected by \textit{INTEGRAL} on 2004 April 22 at 06:58:02 UTC (Apr 22.290) with the IBIS/ISGRI instrument in the 15-200 keV band (\cite{Mere04}). It was localized with coordinates $\alpha=18^{\rm h}42^{\rm m}00^{\rm s}$, $\delta=+01\degr59\arcmin29\arcsec$ (J2000.0) with an uncertainty radius of $2.5\arcmin$. We will thereafter refer to this localization as the IBAS error box. The reported peak flux in the 20-200\,keV range is about 2.7 photons\,cm$^{-2}$\,s$^{-1}$ ($2.5 \times 10^{-7}$ erg\,cm$^{-2}$\,s$^{-1}$). In this paper we present the spectral analysis with the spectrometer SPI and imager IBIS.\\
The counterpart was very promptly searched by ROTSE, which found no new object up to $R\sim16.5$ less than $30\,\rm s$ after the burst (\cite{rotse}), although field crowding creates significant source confusion.
We carried out our search for the afterglow in the $R$, $I$ and $K_{\rm s}$ bands very quickly, within 2 hours after the burst. Lacking a refined X-ray or radio position, the entire IBAS error box was explored. No afterglow candidate was detected in the error box down to the limit $K=14.5$ when compared with the 2\,MASS catalog (\cite{Males04b}). In this paper, we report the discovery of a faint NIR afterglow with a refined off-line analysis. Subsequent observations carried out 64 days after the burst revealed the presence of a likely host galaxy. We present these results in the context of  dark or optically faint bursts and discuss the advantage of rapid follow-up observations in the infrared.

\section{Prompt Gamma ray Emission with \textit{INTEGRAL}}

IBIS and SPI are coded mask instruments
and the photons from a single point source are spread
over all the individual detectors. Spectral extraction is possible
using specifically designed 
software which consists of modelling the illuminated 
mask by a point source of unitary flux placed at
the sky coordinates of the GRB. The model is then
fit to the detected shadowgram in each channel to obtain the rate
and error for each channel.\\
\object{GRB\,040422} was visible in the partially coded field of view ($9.4\degr$
off-axis) of both IBIS and SPI. The burst was outside the field of view of the two monitoring instruments, JEM-X (Joint European X-Ray Monitor) and OMC (Optical Monitoring Camera).
 This GRB falls into
the class of long bursts with a duration $T_{90}=4\,\rm s$
obtained from the IBIS light curve. The GRB was localized at $\alpha=18^{\rm h}42^{\rm m}01\fs2$,
$\delta=+1\degr59\arcmin01\farcs6$ (J2000.0) in the IBIS data with an error radius of $1.7\arcmin$. Note that this localization was obtained by an off-line analysis that was not available when we triggered our VLT observations, so that it was not used for the search of the afterglow. It is nevertheless fully consistent with the IBAS error box of $2.5\arcmin$ we used for this purpose. The position of the burst, extracted from the SPI data in the energy range 20\,-\,100\,keV with S/N=11.7, is $\alpha=18^{\rm h}42^{\rm m}00\fs7$, $\delta=+1\degr53\arcmin13\farcs2$ (J2000.0) which is $6\arcmin$ away from the IBIS location, consistent with the $10\arcmin$ location accuracy of SPI for a source with S/N $\sim$ 10.

\begin{figure}
\centering
\includegraphics[width=0.7\linewidth, angle=270]{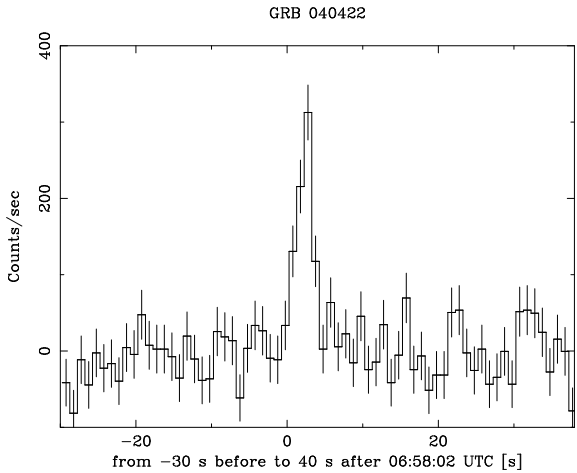}
\caption{Light curve of \object{GRB\,040422} relative to the IBAS trigger extracted from SPI
  with timing resolution of 1 second from 20~keV to 8~MeV.}
\label{spi_lc}
\end{figure}


\begin{figure}
\centering
\includegraphics[width=1.0\linewidth]{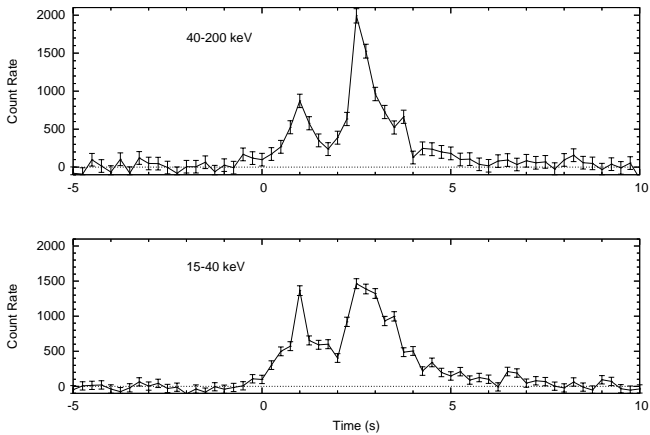}
\caption{Light curves of \object{GRB\,040422} extracted from IBIS/ISGRI data in the energy
ranges 15-40~keV (top) and 40-200~keV (bottom) with 0.25\,s timing resolution.}
\label{ibis_lc}
\end{figure}

\begin{figure}
\centering
\includegraphics[width=0.7\linewidth, angle=270]{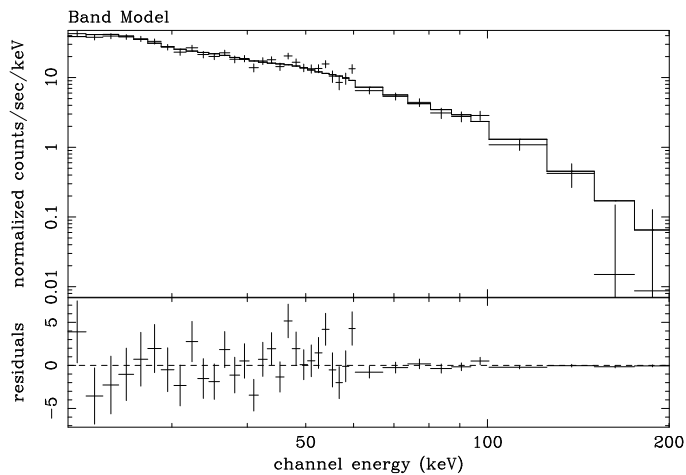}
\caption{IBIS/ISGRI spectrum of \object{GRB\,040422} fitted by a Band model from 20-200 keV. Upper panel: data and best fit model. Lower panel: residuals between the data and the folded model.}
\label{ibis_sp}
\end{figure}

\subsection{SPI analysis}

The determination of SPI light curves with 
 binning on a short time scale is possible
only using the $1\,\rm s$ count rates of the germanium detectors, which 
are usually used for scientific house-keeping purposes. These values 
reflect the count rates of each detector 
in the broad SPI energy band from 20~keV to 8~MeV. The SPI light curve in 
Fig.~\ref{spi_lc} was generated by summing the background-subtracted rates. The background 
was determined from a 40 minute period before the burst trigger and was 
subtracted from each detector individually. The SPI light curve is a single
pulse that does not have the FRED profile that is
observed in many singled-peaked bursts, approximately 7\% of BATSE bursts (\cite{fishman1995}).\\
A spectrum was extracted for one time interval of 4~s starting 
at 06:58:02~UTC.
All single events detected by SPI, corrected for
intrinsic deadtimes and telemetry gaps, were binned into 5 equally spaced
logarithmic energy bins in the 20\,-\,200\,keV range for the 4 second duration
of the burst. 
Version 4.1 of the OSA from the \textit{INTEGRAL} Science Data Centre
(\cite{couv2003}) and the software package SPIROS 8 (\cite{skin2003}) were used
for SPI spectral extraction, while 
XSPEC~11.2 was used for model fitting. 
The background was determined from a 40\,minute period of SPI data before the burst trigger as for the light curve. 
The
spectrum obtained is best fit by a single power--law
model yielding a photon index $\Gamma=-2.17^{+0.24}_{-0.28}$ in the energy range
20--200~keV, with a reduced $\chi^2$ of 2.5 with 3 degrees of 
freedom and the errors are quoted for 1 parameter of interest at the 90\,\%
confidence level. The peak flux obtained in the same energy range is
2.8~photons\,cm$^{-2}$\,s$^{-1}$. 
The same analysis was performed for multiple events incident on the SPI detectors, but yielded no significant improvement to the fit.
The Band model (\cite{band1993}) was also fit to the SPI data for GRB\,040422 yielding a low value for the break energy $E_0 = 60\pm 21\,\rm keV$. However
there is a low probability (0.44,
\textit{f-test}) that the addition of extra model components improves the fit
over a single power--law.

\subsection{IBIS analysis}
\label{secibis}

The data were obtained  from the ISGRI detector of IBIS, an array of 128 x 128 
CdTe crystals sensitive to lower-energy gamma-rays (\cite{leb:2003}), 
 was used to produce the 
light curves of \object{GRB\,040422}. 
The IBIS time profile of the GRB is given in Fig.~\ref{ibis_lc} for the energy
ranges 15-40 and 40-200\,keV with a time resolution of 0.25\,s. On this
time scale, the GRB is resolved into two pulses and the second pulse
is harder and more intense than the first.
A spectrum was extracted for one time interval of 4~s starting 
at 06:58:02~UTC.
The spectrum was fit by
a single power--law model with a photon index of $\Gamma=-2.2\pm0.4$
with a reduced $\chi^2$ of 2.4 for 36 degrees of freedom,
consistent with the SPI determination. 
The Band model was a better fit to the IBIS data yielding values
of the break energy, $E_0=56\pm2\,\rm keV$; 
photon index below the turnover $\alpha=-1.26\pm0.03$ (hence the peak energy is $E_{\rm peak}\equiv(\alpha+2)E_0=41\pm3\,\rm keV$);
and photon index above the turnover $\beta$ fixed at $-$4.0 
 where the errors quoted are for 1 parameter of interest at the 90\,\%
confidence level (Fig.~\ref{ibis_sp}). This fit
resulted in a reduced $\chi^2$ of 1.14 for 35 degrees of freedom. 
The break energy, $E_0$, is in the middle of the energy
band where there is good sensitivity. The peak flux obtained for the Band model in the range 20Ð-200~keV is $2.3^{+0.3}_{-0.5}\,\rm photons\,cm^{-2}\,s^{-1}$ ( $1.8^{+0.3}_{-0.6}\times 10^{-7}\,\rm erg\,cm^{-2}\,s^{-1}$) and the fluence in the same energy range is $3.4^{+1.3}_{-1.0}\times 10^{-7}\,\rm erg\,cm^{-2}$.

\section{Observations with the VLT}
\label{observations}

\begin{table*}
\centering\caption{Log of the observation for \object{GRB\,040422}. Observing time refer to the middle of exposures. Upper limits refer to a non detection at $1\,\sigma$ level with a confidence level of 95\%.}
\label{tab:obslog040422}
\begin{tabular}{lllllll}
\hline
\bf Date         & Time &Seeing FWHM&Number$\times$Exposure  time      & Instrument   & Filter   &Magnitude      \\
(UTC)             &since burst   &$\arcsec$  &      &       &\\
\hline          
2004 Apr 22.37   & 1.90 hours       &0.9        &5$\times$30\,s        &VLT+ISAAC     &$K_{\rm s}$  & $18.0\pm0.1$        \\
2004 Apr 22.37   & 1.93 hours       &0.9      &1$\times$120\,s             &VLT+FORS\,2   &$R$       & $> 24.2$     \\
2004 Apr 22.38   & 2.09 hours       &1.1           &1$\times$120\,s             &VLT+FORS\,2   &$I$         & $> 23.4$ \\
2004 May 05.33   & 13.04  days    &1.8           &5$\times$30\,s        &VLT+ISAAC     &$K_{\rm s}$  & $>   20.2$        \\
2004 Jun 26.11   & 64.82  days    &0.8      &30$\times$60\,s        &VLT+ISAAC     &$K_{\rm s}$  & $20.3\pm0.2$        \\               
\hline
\end{tabular}
\end{table*}

\begin{figure}
\centering
\resizebox{\hsize}{!}{\includegraphics[width=17cm]{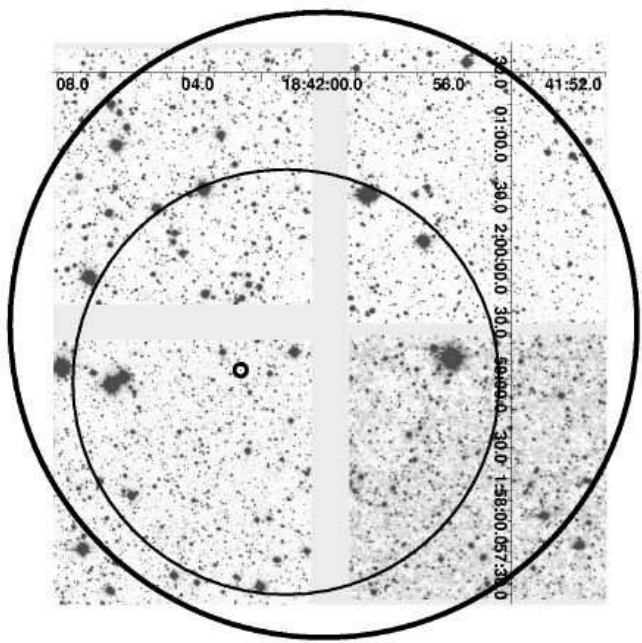}}
\caption{A mosaic of the four panels observed by ISAAC. The large circle corresponds to the IBAS error box, the smaller thin one to the IBIS localization. The smallest circle is centred on our identification of the afterglow. North is up, East is left.}
\label{fig:mosaic}
\end{figure}

Table~\ref{tab:obslog040422} shows the log of observations. Due to the size of the error box compared with the ISAAC field of view ($2.5\arcmin\times2.5\arcmin$), a mosaic of four panels was necessary in the NIR. In the optical, however, just one exposure was sufficient to cover the error box. The raw frames were processed using the standard procedures: subtraction of an averaged bias frame, division by a normalised flat frame, and, for the $K_{\rm s}$ frames, subtraction of a sky model obtained by taking the median of the dithered frames, followed by
positionally registering the frames before co-adding them. Astrometry was performed using the GSC-2 catalog\footnote{\texttt{http://www-gsss.stsci.edu/gsc/gsc2/GSC2home.htm}} for the FORS\,2 frames, and the 2\,MASS\footnote{\texttt{http://irsa.ipac.caltech.edu/}} for the ISAAC frames. In the latter case, about 100 stars per panel were available, leading to a rms of $0\farcs20$. The photometric measurements were made with SExtractor (\cite{bertin}) for the full catalogs, and with \textit{IRAF}\footnote{\textit{IRAF} is distributed by the National Optical Astronomy Observatories,
    which are operated by the Association of Universities for Research
    in Astronomy, Inc., under cooperative agreement with the National
    Science Foundation.} daophot task for single objects of interest. The calibration was done against the standard field \object{CS\,62} (\cite{graham}) for FORS\,2, and against the standard star \object{S808$-$C} (\cite{persson}) for ISAAC. In the NIR, a  $\sim -0.2$ mag correction was made for consistency with the 2\,MASS catalog, using all the available stars of this catalog with $10\leq K_{\rm s}\leq 15$ ($\sim50$ stars per panel) and checking that the results are robust in respect with this choice of magnitude range.\\
A mosaic of the four panels is shown in Fig.~\ref{fig:mosaic}, covering $\sim81\%$ of the IBAS error box and showing $\sim6\cdot10^3$ objects within it. 
We used for comparison the frames of the second epoch, which is significantly deeper than the 2\,MASS catalog. We will assume throughout the paper that the afterglow light curve can be simply modelled by $F(t,\nu)\propto t^{-\alpha}\nu^{-\beta}$ (e.~g. \cite{paradjis}): assuming a temporal index $\alpha$ of 1 with no jet break, we expect a magnitude difference of about 5.5 between the first and second epoch. As the seeing was poorer during our observations in second epoch, we degraded accordingly the images of the first epoch (with psfmatch task of {\it IRAF\/}). 
After aligning the first and second epoch frames, we derived with the SExtractor software the list of detected sources at $2\,\sigma$ and $1\,\sigma$ confidence levels for the first and second epoch, respectively.\\
We set the following criterion to identify the potential afterglow 
candidates:
\begin{description}
\item[1)] the object is seen in the first epoch, its FWHM is larger than $0\farcs7$, its ellipticity is lower than 0.5 (as we are looking for a stellar-like object)~, and: 
\item[2a)] no object is seen within a $0\farcs3$ radius in the second epoch (i.~e. $1.5\,\sigma$ of the astrometric error), or
\item[2b)] an object is seen at the right position, but is at least 2 magnitude fainter (that would correspond to a temporal index of 0.36, an unusually low value).
\end{description}
The number of potential candidates after this step is about 60 for each panel. The choice of different thresholds for the first and second epochs has helped to lower the number of false detections. We subsequently permormed a careful visual inspection to eliminate spurious candidates:
\begin{itemize}
\item cosmic rays;
\item double stars separated in the first epoch, but blended in the second epoch;
\item objects blended with a close bright object;
\item objects close to a bad pixel region;
\item diffuse objects;
\item objects that can be glimpsed in the second epoch, but escaped detection because their fluxes were below threshold.
\end{itemize}
Only one candidate remains after this careful inspection. It is isolated and stars with similar magnitudes are well above the detection threshold on the reference frame. Its coordinates are: $\alpha=18^{\rm h}42^{\rm m}02\fs65$, $\delta=01\degr59\arcmin07\farcs35$ (J2000.0), i.~e. $45\arcsec$ off the centre of the IBAS error box, and $22\farcs5$ off the centre of the IBIS error box reported in Sect.~\ref{secibis}. Its $K_{\rm s}$ magnitude at the first epoch is $K_{\rm s}=18.0\pm0.1$, as reported in Table~\ref{tab:obslog040422}. This candidate is not detected in the second epoch frame. 
Careful analysis of the $R$ and $I$ frames yielded only upper limits given in Table~\ref{tab:obslog040422}.\\
In order to confirm the candidate afterglow we made a third set of observations with ISAAC about two months after the GRB, with a much longer integration time. This observation revealed, at the position of the candidate, a faint object with an ellipticity of $\sim0.7$, which spans $\sim2.5\arcsec$ in its largest dimension. Its magnitude is $K_{\rm s}=20.3\pm0.2$. This corresponds to a $2\,\sigma$ detection, the $1\,\sigma$ detection limit being $K_{s}=21$ (95\% C.L.). Over the $\sim 400$ objects detected in the range $20< K_{\rm s}<21$, only 0.7\% have a greater ellipticity, indicating that it is unlikely that this object is a variable star. The possibility of a QSO has also been considered. The number of QSOs with $B<21$ is about 100 per square degree (\cite{ulrich}), hence we can expect 0.5 in the IBAS error box. However, only a small fraction of these objects, the Optically Violently Variable (OVVs, or BL Lac), may have a variation of more than 2 mag in 13 days, to comply  with our observations in the first and second epochs. The amplitude of variability is expected to depend only weakly on the wavelength. The OVVs are not objects that have bursts and become very faint between two bursts, they have an erratic behaviour between two extrema that can reach 5 mag or more.
An estimate of the number of these objects is given by \cite{collinge}, about 0.15 per square degree. The probability to find by chance such an object in the IBAS error box is therefore very small, about 0.1\%. Then, although a firm rejection of this hypothesis cannot be made because of the scarcity of the data, it seems very unlikely that the object is an OVV.
We therefore suggest that this object is the host galaxy of \object{GRB\,040422}. 
This deep observation strengthens our identification of the afterglow of \object{GRB\,040422}.\\
\begin{figure}
\centering
\resizebox{\hsize}{!}{\includegraphics[width=17cm]{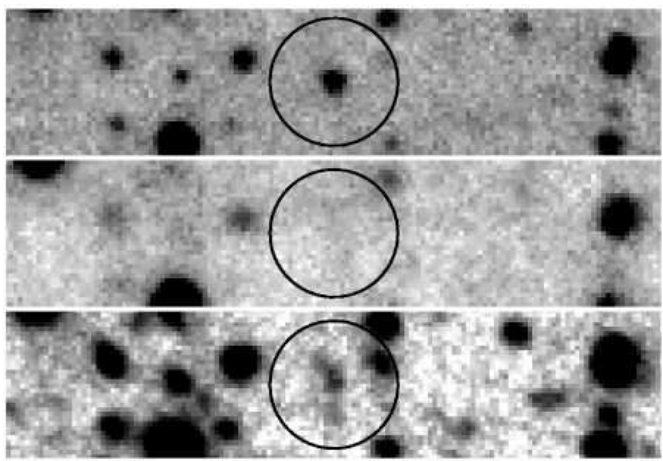}}
\caption{The region of the candidate afterglow. From top to bottom: 2004 Apr 22.37, 2004 May 05.33, 2004 Jun 26.11. The circle around the afterglow on the first epoch has a radius of $2\arcsec$ to give the scale. The second epoch is affected by bad seeing. A logarithmic scale was chosen for the third epoch to bring out the elongated shape of the likely host galaxy.}
\label{fig:epoch}
\end{figure}
\begin{figure}
\centering
\resizebox{\hsize}{!}{\includegraphics[width=17cm]{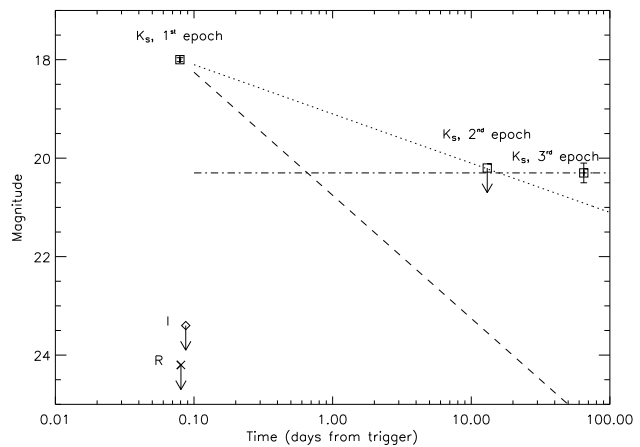}}
\caption{The light curve of the afterglow of \object{GRB\,040422}. Two power-law models are included,  with temporal indices of $\alpha=0.4$ (dotted line) and $\alpha=1$ (dashed line). The horizontal line gives the magnitude of the host.}
\label{fig:light curve}
\end{figure}
A close-up around the afterglow position for the three epochs is shown in Fig.~\ref{fig:epoch}. The light curve is shown in Fig.~\ref{fig:light curve}. Using the first two epochs, we can constrain the temporal index to be $\alpha>0.4$, but this could be an average between two regimes in case of a jet break. 
This is consistent with the decay of observed afterglows, which have $\alpha$ in the range $0.7 - 2$ (\cite{paradjis}). Assuming a typical decay with $\alpha = 1$, the contribution of the afterglow at the third epoch is negligible. Also, any associated supernova is not expected to contribute significantly at this late time. Conversely, the contribution of the host galaxy to the afterglow flux during the first epoch is less than 0.2 mag. We can also note that the upper limit of ROTSE (\cite{rotse}) would translate to $R>22.4$ in our first epoch for $\alpha=1$, or that the ROTSE limit and our limit are equivalent within the power-law framework for $\alpha=1.3$. Such values for $\alpha$ are within the usual range.\\
The magnitudes reported in Table~\ref{tab:obslog040422} are the observed values, and do not take into account the absorption due to the Milky Way. The afterglow lies towards the Galactic plane, rather close to the direction of the Galactic centre ($l=33\degr37\arcmin51\farcs85$, $b=2\degr59\arcmin41\farcs83$). So close to the Galactic plane, the extinction is high, rather patchy and uncertain. The hydrogen column density is estimated to be $N_{\rm H}=6\times10^{21}\,\rm cm^{-2}$  (\cite{dickey}). Using the fit of \cite{predehl}, we derive $A_{V}=3.6$, and hence, using the extinction law of \cite{cardelli} with $R_{V}=3.1$, we get $E(B-V)_{N_{\rm H}}=1.17$. The use of the dust emission maps at $100\,\mu\rm m$  by FIRAS obtained by \cite{schlegel} leads to a substantially higher absorption of $E(B-V)_{\rm FIR}=1.92$, although according to the authors, the calculations should not be trusted for Galactic latitudes below $5\degr$. This overestimate of the reddening has been considered by \cite{dutra}, who proposed a linear correction: $E(B-V)=0.748E(B-V)_{\rm FIR}+0.0056$, valid for $0.7<E(B-V)<1.6$. In our case, we get $E(B-V)=1.43$ (hence in the validity domain of the relation), and then $A_V=4.4$, $A_R=3.8$, $A_I=2.8$ and $A_K=0.5$. In the rest of the paper, we will keep these values for definiteness.
As expected, infrared observations are mandatory to look for afterglows within the Galactic plane, in which the majority of the \textit{INTEGRAL} alerts are concentrated. We can use these absorption values to constrain the power-law spectral index $\beta$ using our $R$ and $I$ upper limits combined with our $K_{\rm s}$ measurement (first epoch). As the frames were acquired almost simultaneously, the temporal evolution has little effect. We get $\beta>1.7$ (conversely, if we suppose the more common value of $\beta=0.7$, we derive $E(B-V)>2.96$).  
This rather high value for $\beta$ might indicate the presence of an additional, intrinsic reddening, or that there is Lyman-$\alpha$ damping, although the safer explanation is that our estimate of the Milky Way absorption is inaccurate. We remark anyway that an afterglow with such a steep spectral index goes naturally undetected in the $R$ band where most searches are done, while it is much easily detected in the $K_{\rm s}$ band. In other words, the observed distribution of $\beta$ is biased towards low values.

\section{Discussion}

\subsection{The prompt Gamma ray emission: comparison with other \textit{INTEGRAL} bursts}

Twenty GRBs occurred in the field of view of
the main \textit{INTEGRAL} instruments up to the end of 2004.
One of these events, initially classified as a GRB  
(\object{GRB\,040903}) is believed to be an X--ray flash (XRF) or 
a possible Type I X--ray flare from a new 
transient source in the Galactic bulge (\cite{gmpm04}). 
In a second case, \object{GRB\,031203}, 
modelling of the dust-scattered X--ray echo provided the first 
evidence for a low luminosity, XRF source (\cite{vwo+04}; \cite{whl+04}).
The IBIS spectrum, however, is consistent with a single power--law 
of photon index $-1.63\pm0.06$ (\cite{sls04}; \cite{soderberg2004}), typical of \textit{INTEGRAL} 
bursts. This event is the only \textit{INTEGRAL} burst to date for which a 
direct redshift ($z=0.1055\pm0.0001$, \cite{pro}) measurement has been made.\\
A single power--law model, with photon index in the
range $-1$ to $-2$, provides a good fit to the data 
for the vast majority of \textit{INTEGRAL} bursts in 
the range 20 to 200~keV (e.~g. \cite{vonK2003}, \cite{moran}; \cite{mere}).  However, 
the SPI spectral index 
 of \object{GRB\,040422} is very steep and lies outside this range 
for a power--law fit with $\Gamma=-2.17^{+0.24}_{-0.28}$. 
Prior to this burst the steepest photon index for an \textit{INTEGRAL}
burst was $\Gamma=-1.90\pm0.15$ for \object{GRB\,040403} (\cite{mere}).
In addition, prior to \object{GRB\,040422}, \object{GRB\,030131} was the only burst
that was best fit by a Band model over its whole 
duration, with a break energy, $E_0$, of $70\pm20\,\rm keV$, a 
photon index below the turnover, $\alpha$, of
$-1.4\pm0.2$ and photon index above the turnover, $\beta$, of $-3.0\pm1.0$
 (\cite{gotz2003}). \object{GRB\,040422} and \object{GRB\,030131} have similar spectral
properties.
The temporal properties of a large sample of bursts detected by the
BATSE experiment are available
 (\cite{quilligan:2002}; \cite{mcbreen:2002}) and \object{GRB\,040422} is consistent with this
sample.  
There is nothing to indicate that this burst is anything other 
than a normal cosmological GRB.

\subsection{The afterglow and the host galaxy}

\begin{table*}
\centering\caption{A compilation of afterglow data. References are given in superscripts.  Note that the magnitude measurements are not necessarily the first available: in some cases, later but more accurate measurements have been preferred.}
\label{tab:afterglow}
\begin{tabular}{lllllll}
\hline
Name&Redshift&Temporal index&Spectral index&Magnitude&Time (hours)&Host galaxy\\
\hline
\object{GRB\,970228}&     $0.695^{\rm (1)}$&     $1.10\pm0.1^{\rm (2)}$&    0.61$\pm$0.32&$R=20.9^{\rm (3)}$&      20&$R=24.6$, $K=22.6^{\rm (4)}$\\
\object{GRB\,971214}&      $3.418^{\rm (5)}$&     $1.20\pm0.02^{\rm (6)}$&    $0.93\pm0.06^{\rm (7)}$&$K=18.03^{\rm (8)}$&      3.5&$R=25.6$, $K=22.4^{\rm (4)}$\\
\object{GRB\,980326}&   $1.0^{\rm (9)}$   &     $2.\pm0.1^{\rm (9)}$&    $0.8\pm0.4^{\rm (9)}$&$R=21.25^{\rm (9)}$&      11.1&$V=29.3^{\rm (10)}$\\
\object{GRB\,980519}&      &     $2.30\pm0.12^{\rm (11)}$&  $1.4\pm0.3^{\rm (11)}$&$R=20.28^{\rm (11)}$&      15.59&$R=26.05^{\rm (12)}$\\
\object{GRB\,990123}&     $1.598^{\rm (13)}$&     $1.10\pm0.03^{\rm (14)}$&    $0.8\pm0.1^{\rm (14)}$&$R=18.65^{\rm (14)}$&      4.06&$R=24.3$, $K=21.9^{\rm (4)}$\\
\object{GRB\,990510}&     $1.619^{\rm (15)}$&    $0.76\pm 0.01^{\rm (16)}$& $0.61\pm0.12^{\rm (16)}$&$R=17.54^{\rm (17)}$&      3.45&$R=27.5^{\rm (4)}$\\
\object{GRB\,010921}&    $0.45^{\rm (18)}$&$1.59\pm0.18^{\rm (18)}$& $2.22\pm0.23^{\rm (18)}$&$R=19.4^{\rm (19)}$&      21.8&$R=21.45$, $K=19.05^{\rm (4)}$\\
\object{GRB\,011121}& $0.36^{\rm (20)}$&$1.72\pm0.05^{\rm (20)}$&$0.66\pm0.13^{\rm (20)}$&$R=19.06^{\rm (20)}$&      10.37&$R=24.6^{\rm (4)}$\\
\object{GRB\,020405}&$0.69^{\rm (21)}$&     $1.54\pm0.06^{\rm (21)}$&     $1.3\pm0.2^{\rm (21)}$&$R=20.17^{\rm (21)}$&      23.62&$R=20.9^{\rm (4)}$\\
\object{GRB\,020813}&$1.255^{\rm (22)}$& $0.76\pm0.05^{\rm (23)}$&    $1.04\pm0.03^{\rm (23)}$&$R=18.49^{\rm (23)}$&      3.97&$R=24.7^{\rm (4)}$\\
\object{GRB\,021211}&$1.004^{\rm (24)}$&$1.11\pm0.01^{\rm (25)}$&$0.6\pm0.2^{\rm (25)}$&$R=21.9^{\rm (25)}$&6.81&$R=25.16^{\rm (25)}$\\
\object{GRB\,030329}&     $0.1685^{\rm (26)}$&    $0.89\pm    0.01^{\rm (27)}$&     $0.71^{\rm (28)}$&$R=15.02^{\rm (28)}$&      9.08&\\
\object{GRB\,030429}&      $2.658^{\rm (29)}$&    $0.95\pm0.03^{\rm (29)}$&    $0.36\pm0.12^{\rm (29)}$&$K=17.70^{\rm (29)}$&      13.63&\\
\object{GRB\,030528}&&$\sim1.2^{\rm (30)}$&&$K=18.6^{\rm (30)}$&16.6&$R=22.0^{\rm (30)}$\\
\object{GRB\,031203} & $0.1055^{\rm (31)}$&$\gse2^{\rm (32)}$&$2.36\pm0.02^{\rm (32)}$&$K=17.56^{\rm (32)}$&9&$K=16.5^{\rm (32)}$\\
\object{GRB\,040924}&$0.859^{\rm (33)}$&$0.7^{\rm (34)}$&  &$K=17.5^{\rm (35)}$&      2.40&$K=20.4^{\rm (36)}$\\
\object{GRB\,041006}&  $0.712^{\rm (37)}$    &$0.7^{\rm (38)}$&$0.45^{\rm (39)}$&$R=18.5^{\rm (38)}$&      1.90&\\
\hline
\end{tabular}
\begin{flushleft}
$^{\rm (1)}$\cite{blo},
$^{\rm (2)}$\cite{fru},
$^{\rm (3)}$\cite{mas},
$^{\rm (4)}$\cite{lefloch} and references therein,
$^{\rm (5)}$\cite{kulk},
$^{\rm (6)}$\cite{die},
$^{\rm (7)}$\cite{rei},
$^{\rm (8)}$\cite{goro},
$^{\rm (9)}$\cite{blo2},
$^{\rm (10)}$\cite{chary},
$^{\rm (11)}$\cite{vrba},
$^{\rm (12)}$\cite{sok},
$^{\rm (13)}$\cite{hj},
$^{\rm (14)}$\cite{kulkarni},
$^{\rm (15)}$\cite{vre},
$^{\rm (16)}$\cite{stanek},
$^{\rm (17)}$\cite{harrison},
$^{\rm (18)}$\cite{pri},
$^{\rm (19)}$\cite{par},
$^{\rm (20)}$\cite{gar},
$^{\rm (21)}$\cite{mas2},
$^{\rm (22)}$\cite{barth},
$^{\rm (23)}$\cite{cov},
$^{\rm (24)}$\cite{valle},
$^{\rm (25)}$\cite{pandey},
$^{\rm (26)}$\cite{gre},
$^{\rm (27)}$\cite{torii},
$^{\rm (28)}$\cite{math},
$^{\rm (29)}$\cite{jakobsson},
$^{\rm (30)}$\cite{rau},
$^{\rm (31)}$\cite{pro},
$^{\rm (32)}$\cite{malesani031203},
$^{\rm (33)}$\cite{wie},
$^{\rm (34)}$\cite{fox},
$^{\rm (35)}$\cite{terada},
$^{\rm (36)}$from \cite{teradahost}, 27.7 hours after the burst, includes a contribution from the afterglow, 
$^{\rm (37)}$\cite{fug},
$^{\rm (38)}$\cite{pri2},
$^{\rm (39)}$computed from \cite{costa}.
\end{flushleft}
\end{table*}

The data that we presented here are the only ones available to date for the afterglow of \object{GRB\,040422} with a positive detection. They are rather scarce, the main reason being that, because of the localization of the burst within the Galactic plane, the afterglow had not been localized quickly enough to undertake variability and spectroscopic studies. 
It is interesting to note that the host galaxy ($K_{\rm s}=19.8\pm0.2$ after correction for the Milky Way reddening) is rather bright for a GRB host galaxy: in the sample of 19 host galaxies of \cite{lefloch} (with $0.43\leq z\leq 3.42$ and $19.05\leq K\leq23.5$), only two are brighter: the host of \object{GRB\,010921}, with $K_{\rm s}=19.05\pm0.1$ and $z=0.45$, and the host of \object{GRB\,980703}, with $K_{\rm s}=19.6\pm0.1$ and $z=0.97$. This might suggest that the galaxy has a rather small redshift. However, the strong Galactic absorption will prevent observations in the optical, and photometric observations in $J$, $H$, $K_{\rm s}$ will probably be insufficient for a photometric redshift if we face a starbust galaxy. Infrared spectroscopy is difficult and time consuming, even for the VLT: about a whole night would be needed for ISAAC to get a sufficient signal to noise ratio. We are facing a case where an early spectrum of the GRB afterglow would have been possibly the only way to securely derive the redshift.\\   
We compared the afterglow of \object{GRB\,040422} with those of GRBs listed in Table~\ref{tab:afterglow} selected according to the following criteria:
\begin{itemize}
\item a photometric measurement of the optical counterpart have been made in the first 24 hours with good observing conditions; this timing constraint has two motivations: first, to avoid a bias towards bright afterglows that late detections tend to select; second, to avoid introducing a large error in extrapolating to the first epoch of our observations; 
\item the temporal index $\alpha$ is available to extrapolate to the first epoch of our observations of \object{GRB\,040422} assuming that the lightcurve can be modelled by $F(t,\nu)\propto t^{-\alpha}\nu^{-\beta}$; in case of a temporal break, only the index before the break is adopted;
\item the $K_{\rm s}$ magnitude is given, or the spectral index $\beta$ is given so that the extrapolation to this band can be made.
\end{itemize}
\begin{figure}
\centering
\resizebox{\hsize}{!}{\includegraphics[width=17cm]{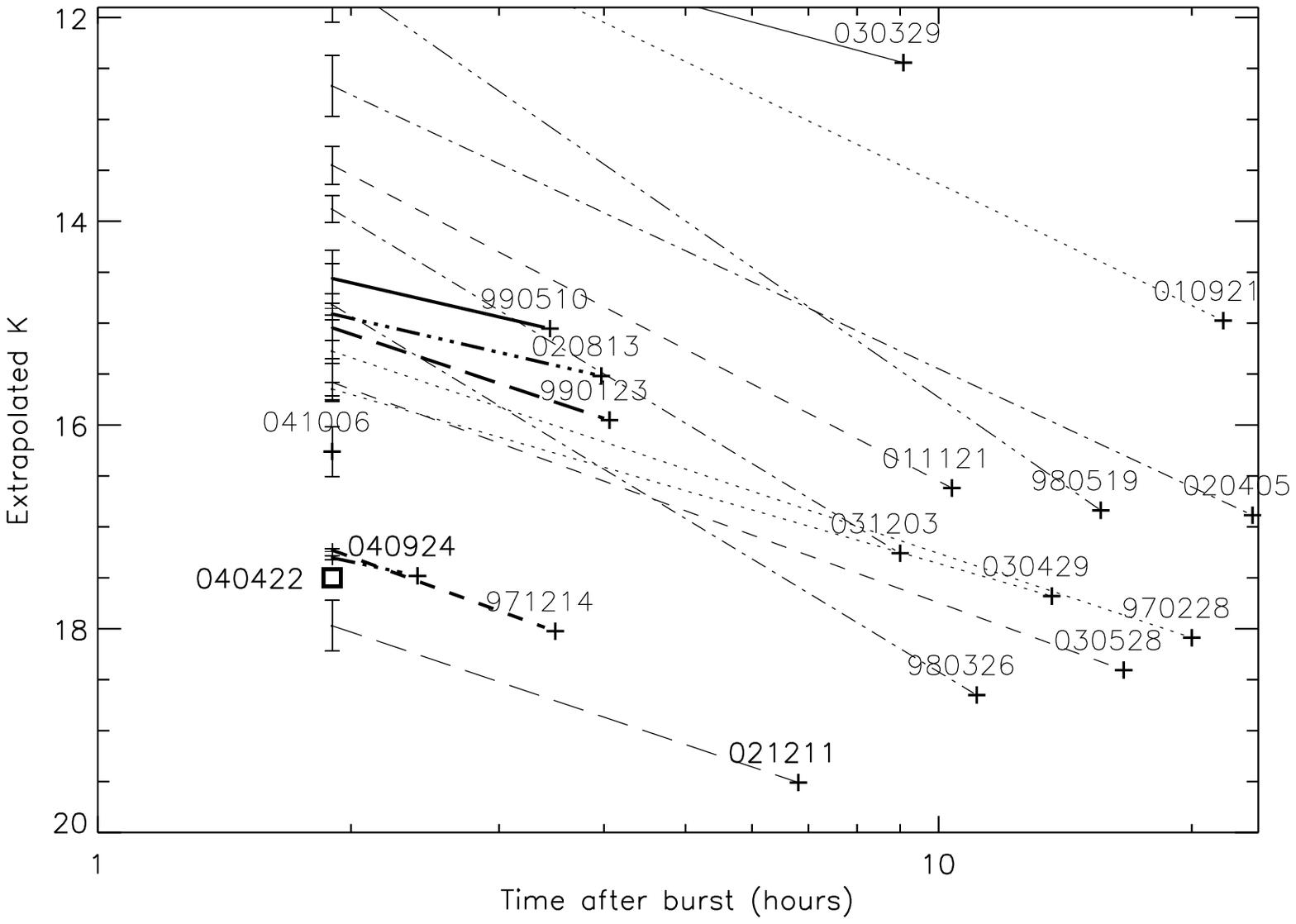}}
\caption{Light curves of the afterglows listed in Table~\ref{tab:afterglow}, with magnitude extrapolated to the $K_{\rm s}$ band after correction of the Milky Way absorption. The cross is placed at the time of observation. The lines indicate the temporal power-law decay only backward in time up to the the time of observation of \object{GRB\,040422} for clarity, different line styles are used for clarity only. Our measurement of the afterglow of \object{GRB\,040422} is indicated by a square. The error bars indicate the uncertainty of the extrapolated $K_{\rm s}$ magnitude at the time of our observations given the uncertainties on the indices given in Table~\ref{tab:afterglow} and assuming that the power law model is valid. The extrapolated lightcurve of afterglows observed less than six hours after the bursts are marked in bold.
}
\label{fig:dark}
\end{figure}
Using the spectral index, we extrapolate their magnitude to the $K_{\rm s}$ band and correct for the absorption in the Milky Way (negligible for our purpose except for \object{GRB\,031203}), using, as for \object{GRB\,040422}, the $E(B-V)$ values from the FIRAS maps corrected according to \cite{dutra}. We then extrapolate to the epoch of detection of \object{GRB\,040422} (1.90 hours) with the reported temporal index. The result is presented in Fig.~\ref{fig:dark}. The afterglow of \object{GRB\,040422} is shown, after correcting for the Milky Way absorption. The error bars indicate the uncertainty of the extrapolated $K_{\rm s}$ magnitude at the time of our observation given the uncertainties in the indices given in Table~\ref{tab:afterglow}, assuming that the power law model is valid (when there is no error reported in Table~\ref{tab:afterglow}, we assume an error of 0.08 for the temporal index, 0.2 for the spectral index, i.~e. the average of the errors listed in the table).\\
A cautionary comment must be made at this point. We choose a simple modelling for the extrapolation, which could introduce biases in mainly two ways.
\begin{itemize}
\item The temporal curve may deviate from a power law behaviour at early times: e.~g. \object{GRB\,970508} was nearly constant during the first day before decaying (\cite{djorgovski}), while \object{GRB\,990123} showed an optical flash was detected in the first minutes (\cite{akerlof}). Moreover, afterglows tend to have steeper temporal indices at later times (e.~g. \cite{foxtemp}). This is apparent in Figure~\ref{fig:dark}, where we have marked in bold the extrapolated lightcurves of afterglows observed less than six hours after the burts (i.~e. three times the delay of our observation of \object{GRB\,040422}), for which we can expect that a lower extrapolation error and which are therefore more significant for our purpose.   
\item The absorption induces variations with wavelength of the spectral index. This can lead to an overestimate of the $K_{\rm s}$ flux extrapolated from the observed $R$ magnitude that is more affected by extinction. Interstingly, in Figure~\ref{fig:dark}, six of the nine faintest points are $K_{\rm s}$ measurements, and all the nine brightest points are $R$ measurements extrapolated to the $K_{\rm s}$ band. This suggests that extinction might be playing a role for some of these GRBs. As already discussed, the extinction due to the Milky Way is subject to inaccuracy, especially at low Galactic latitudes. However, the afterglow of \object{GRB\,011121} is the only case for which we made an extrapolation to the $K$ band with $A_R(\rm Milky\;Way)>1$. For \object{GRB\,021211}, we have only $A_R(\rm Milky\;Way)=0.067$. The estimation of the extinction within the host galaxy is beyond the scope of this paper. However, the study of \cite{stratta} indicates that the reddening might be low: in the rest frame of the host, $A_V(\rm host)=0.21\pm 0.12$ for \object{GRB\,971214} and is consistent with zero for \object{GRB\,980519}, \object{GRB\,990123} and \object{GRB\,990510}. For \object{GRB\,020813}, \cite{savaglio} infer a relatively high extinction of $A_V(\rm host)\simeq 0.4$, but with a weak dependance on wavelength, and therefore little impact on the power law shape of the spectrum. This suggests that, at least for the afterglows of Table~\ref{tab:afterglow} observed in the $R$ band, the extinction within the host might be small.
\end{itemize}
Keeping these caveats in mind, Figure~\ref{fig:dark} shows that the afterglow of \object{GRB\,040422} appears to be significantly dimmer than most of those shown in Fig.~\ref{fig:dark}. 
The only comparable afterglows are the one of \object{GRB\,021211}, already discussed by \cite{pandey} in the context of dark bursts and qualified as ``not so dark'', and that of \object{GRB\,040924}, observed 2.4 hours after the burst in the $K$ band by the Subaru telescope by \cite{terada} following the promptly localization in the $R$ band by the robotic Palomar 60-inch telescope (\cite{fox040924}). This means that if prompt observations with the VLT have not been triggered, \object{GRB\,040422} would have been classified as dark: indeed, from an observational point of view, it is optically dark and infrared faint.\\
This faintness contrasts with the relative brightness of the host galaxy. Indeed, less than two hours after the burst, it is just 2.3 mag fainter than the afterglow. We give in Table~\ref{tab:afterglow} the magnitudes of the hosts for our sample. Using again the temporal and spectral indices, we extrapolate the magnitude of the afterglow in the band in which the magnitude of the host is given, 1.90 hours after the trigger. We find that on the average the afterglow is 7 mag brighter than its host, this figure still holds if we restrain the sample to the 6 GRBs for which the magnitude of the host is given in the $K$ band. For only two GRBs the difference is close to the one of \object{GRB\,040422} 1.90 hour after the burst: \object{GRB\,040924} (difference 3 mag), and the very extreme case of \object{GRB\,031203} (difference 2.3 mag, actually the afterglow was fainter than the host at the time of observation), which has also a very high spectral index, $\beta=2.37$, to be compared to the high values we derived in Sec.~\ref{observations}. This indicates that \object{GRB\,040422} may share some properties with \object{GRB\,031203} with respect with the environment of the burst and its host galaxy.\\
These results show that 8-m class telescopes operated in the infrared have the capability to explore the dark/faint burst domain, reducing the proportion of burst that are considered dark only because of an inefficient follow-up. Indeed, we may consider the possibility that dark bursts do not really exist, and that their afterglows are simply fainter and with different global spectral properties. Another possibility is that we are observing an afterglow enshrouded by dust surrounding the burst site, although it seems so far that no absorption is the rule. For \object{GRB\,040422}, neither of these possibilities can be excluded. More information could have been gathered by acquiring a spectrum. This would have required a quick localization by a robotic telescope. REM (\cite{zerbi}) has now the capability to promptly  localize such an event: if we assume a temporal index of -1, the afterglow of \object{GRB\,040422} would have $K_{\rm s}=14.6$, 5 minutes after the burst, corresponding to the $5\,\sigma$ detection limit of REM with a 30 second exposure. In 2008, the X-shooter spectrograph (\cite{Moorwood}) will be operational at the VLT and sensitive enough to obtain a spectrum of an afterglow similar to the one of \object{GRB\,040422} in the infrared and in the optical with medium resolution ($5\,000-8\,000$ in the infrared), provided that the observation is as rapid as in the case of \object{GRB\,040422}. With such a spectrum, it will be possible to measure the redshift, to quantify precisely the reddening and to study the properties of the host galaxy in a unique and comprehensive way, increasing our understanding of the GRB environment.

\section{Conclusion}

We have described the results obtained on \object{GRB\,040422} using IBIS 
and SPI on the \textit{INTEGRAL} satellite and the instruments FORS\,2 and 
ISAAC on the VLT. The main results are:
\begin{itemize}
\item the IBIS spectrum is well fitted with the Band model;
\item only two hours after the trigger, the afterglow was below the detection limit of the 2\,MASS catalog, therefore a second epoch observation was required to spot the afterglow;
\item we have detected the afterglow in the $K_{\rm s}$ band, but not in the $R$ and $I$ band, at least partially because of the Milky Way absorption;
\item we have detected the bright host galaxy in the $K_{\rm s}$ band, with $K_{\rm s}=20.3\pm0.2$ to be corrected from a 0.5 mag Milky Way absorption;
\item comparison with a compilation of quickly observed afterglows indicates that the one of \object{GRB\,040422} is the dimmest, with the exception of the afterglow of \object{GRB\,021211} and possibly \object{GRB\,040924}.
\end{itemize}
Without our prompt infrared observations, \object{GRB\,040422} would probably have been classified as dark. It is nevertheless faint in the infrared. This suggests that the proportion of dark GRBs can be significantly lowered by a more systematic use of 8-m class telescopes in the infrared in the very first hours after the burst. Quicker robotic telescopes operating in the infrared, like REM, and the faster and more precise positions that will be provided soon by Swift would of course facilitate the identification, allowing spectrophotometric observations to explore the domain of dim afterglows.

\begin{acknowledgements}
It is a pleasure to thanks to the ESO staff at the VLT for the observations. Data mining for constructing Table~\ref{tab:afterglow} was speeded up by the use of Jochen Greiner's GRB page \texttt{http://www.mpe.mpg.de/$\sim$jcg/grb.html}. We thank D. Lazzati, L. Guzzo, S. Bhargavi and P. V\'{e}ron for useful discussions and comments. PF wishes to thank CNRS/APC for funding and B. Bigot, Haut Commissaire \`{a} l'Energie Atomique, for funding and support. DM wishes to thank the Italian Istituto Nazionale di Astrofisica (INAF).
\end{acknowledgements}

\end{document}